%% file: Camera_Ready.tex
\documentclass[acmsmall]{acmart}

\AtBeginDocument{%
  }

\usepackage{amssymb}
\usepackage{algorithmic}
\usepackage{graphicx}
\usepackage{stfloats}
\usepackage{textcomp}
\usepackage{xcolor}
\usepackage{xspace}
\usepackage[nointegrals]{wasysym}
\usepackage{pifont}
\usepackage{xcolor}
\usepackage{multirow}
\usepackage{flushend}
\usepackage{caption}
\usepackage{booktabs}
\usepackage{hyperref}
\usepackage{algorithm}
\usepackage{algorithmic}
\usepackage{pdfpages}
\usepackage{tabularx}
\usepackage[inkscapelatex=false]{svg}
\hypersetup{
colorlinks = true, 
linkcolor=black, 
citecolor=black, 
urlcolor=black 
}
\usepackage{url}
\usepackage{enumitem}
\usepackage{braket}
\usepackage{amsmath}
\usepackage{caption}
\usepackage[breakable]{tcolorbox}
\usepackage{colortbl}
\usepackage{pifont}
\usepackage{float}                  
\usepackage{subfig}                 
\usepackage{overpic}                

\usepackage{tikz}
\usepackage{color}
\usepackage{listings}
\input{macros}

\setcopyright{acmlicensed}

\begin{document}

\title{Quantum Concolic Testing}

\author{Shangzhou Xia}
\orcid{0009-0006-2775-9633}
\affiliation{%
  \institution{Kyushu University}
  \city{Fukuoka}
  \country{Japan}
}
\email{xiaszore19@gmail.com}

\author{Jianjun Zhao}
\authornote{Corresponding author.}
\orcid{0000-0001-8083-4352}
\affiliation{%
  \institution{Kyushu University}
  \city{Fukuoka}
  \country{Japan}
}
\email{zhao@ait.kyushu-u.ac.jp}

\author{Fuyuan Zhang}
\orcid{0009-0001-6560-5102}
\affiliation{%
  \institution{The University of Tokyo}
  \city{Japan}
  \country{Japan}
}
\email{fuyuanzhang@163.com}

\author{Xiaoyu Guo}
\orcid{0009-0008-4521-9715}
\affiliation{%
  \institution{Kyushu University}
  \city{Fukuoka}
  \country{Japan}
}
\email{guo.xiaoyu.961@s.kyushu-u.ac.jp}

\renewcommand{\shortauthors}{Shangzhou Xia, Jianjun Zhao, Fuyuan Zhang and Xiaoyu Guo}

\begin{abstract}
\textcolor{black}{This paper presents the first concolic testing framework explicitly designed for quantum programs. The framework introduces quantum constraint generation methods for quantum control statements that quantify quantum states and offers a symbolization method for quantum variables. Based on this framework, we generate path constraints for each concrete execution path of a quantum program. These constraints guide the exploration of new paths, with a quantum constraint solver determining outcomes to create novel input samples, thereby enhancing branch coverage. Our framework has been implemented in Python and integrated with Qiskit for practical evaluation. Experimental results show that our concolic testing framework improves branch coverage, generates high-quality quantum input samples, and detects bugs, demonstrating its effectiveness and efficiency in quantum programming and bug detection.}
\textcolor{black}{Regarding branch coverage, our framework achieves more than 74.27\% on quantum programs with under 5 qubits.}
\end{abstract}



\maketitle

\section{Introduction}\label{sec:intro}

With its groundbreaking potential, quantum computing is positioned to transcend the computational boundaries of classical computing systems. Its applications span multiple advanced fields, including artificial intelligence~\cite{dunjko2018machine}, computational chemistry~\cite{mcardle2020quantum,cao2019quantum}, and drug design~\cite{zhou2010quantum,raha2007role}. The evolution of quantum processing unit (QPU) architectures~\cite{wintersperger2022qpu} and the gradual refinement of quantum programming languages~\cite{qiskit,svore2018q,cirq} mark the beginning of a new era of software development tailored for quantum computation. However, the novelty and complexity of quantum programs pose significant challenges in ensuring their reliability and correctness, as highlighted by reports of prevalent bugs in programs written in leading quantum programming languages such as Qiskit~\cite{bugs4q,paltenghi2022bugs,campos2021qbugs}.

\textcolor{black}{The pursuit of robust quantum software has sparked a growing interest in adapting and innovating testing methodologies that are suited to the quantum context. Early attempts to improve the robustness of quantum programs have primarily utilized testing methods~\cite{ali2021assessing,wang2022qusbt, ye2023quratest,shi2024quantest,muskit,wang2021generating,wang2018quanfuzz,long2022testing,long2024testing,long2023equivalence,shao2024coverage} and verification techniques~\cite{kashefi2024verificationquantumcomputationstrusted, bauer2023symqv,fang2023symbolic,Lewis_2023,10.1007/978-3-642-10622-4_7}.
These methods face challenges similar to those encountered in classical software testing, such as test case redundancy and difficulty in fully exploring program behaviors. Verification techniques, in particular, often suffer from high costs and lack flexibility for various programs. The unique quantum properties and limited resources of quantum programs exacerbate these challenges.}

\textcolor{black}{Recognizing the limitations of existing testing methods, this paper introduces a novel approach through the lens of concolic testing~\cite{godefroid2005dart,sen2005cute,majumdar2007hybrid}, a technique often employed in classical software testing for its effectiveness in exploring execution paths and improving the coverage of the test through directed input samples. However, applying concolic testing to quantum programs presents unique challenges. These include the complex representation of quantum variables, which leads to a black-box dilemma, and the lack of constraint representations for the probabilistic outputs inherent in quantum computations, complicating the identification of deterministic execution paths for testing.}

We propose a comprehensive concolic testing framework specifically designed for quantum programs to address these challenges. Central to our framework is the concept of quantum symbolic objects, which enables the symbolic representation of quantum states and makes the black-box nature of quantum computations more transparent. Throughout an execution path, every assignment and quantum operation modifies the program's state using symbolic expressions. We also address the issue of probabilistic outputs by devising constraint formulas for quantum control statements, allowing for a more predictable approach to test case generation. Each conditional statement generates a constraint based on symbolic inputs, and our framework uses these constraints to generate precise test cases that maximize branch coverage. 
\textcolor{black}{Different path constraints lead to varied execution paths, so improving branch coverage can effectively explore program behaviors and detect potential problems.}

To validate the effectiveness of our framework, we conducted a series of experiments to evaluate its performance in improving branch coverage, exploring program branches to detect bugs, and generating high-quality test samples. Our comparative analysis not only benchmarks our approach against existing quantum input case generators but also highlights the nuances of test generation techniques in the quantum realm. Our findings demonstrate the superior performance of our framework in achieving significant improvements in branch coverage, efficiency in program exploration, and the generation of high-quality test samples. Additionally, our framework efficiently finds buggy branches and detects unique quantum bugs that are undetectable by other testing methods. Our approach shows remarkable efficiency, particularly with quantum programs with fewer qubits or reduced program size.

Our contributions in this paper are threefold:

\vspace{-1mm}
\begin{itemize}[leftmargin=1.5em]
\setlength{\itemsep}{2pt}
\item We propose the first concolic testing framework tailored for quantum programs, paving the way for innovative testing strategies in quantum software development.
\item We develop and implement an automated concolic testing framework compatible with Qiskit programs, integrating quantum symbolic objects and four types of quantum constraints into the testing process.
\item Through rigorous experimentation and comparative analysis, we establish the effectiveness and potential of our testing framework, offering insights into its applicability and benefits for future quantum software engineering endeavors.
\end{itemize}
\vspace{-1mm}

The rest of this paper is structured as follows. Section~\ref{sec:background} offers essential background information on quantum programs and concolic testing. Section~\ref{sec:example} provides a motivating example to illustrate our quantum concolic testing. Section~\ref{sec:approach} elaborates on the methodology of our approach. \textcolor{black}{Section~\ref{sec:eval} details the experimental results and the analysis conducted using the Qiskit programs.} Section~\ref{sec:threats} examines potential threats to the validity of our method. Section~\ref{sec:related} reviews work related to our research. Section~\ref{sec:conclusion} summarizes the paper and provides concluding remarks.

\section{Background}\label{sec:background}
\textcolor{black}{This section provides an overview of the foundational concepts and background relevant to quantum programs~\cite{nielsen2010quantum} and concolic testing, which form the basis for the subsequent discussions.}

\subsection{Quantum Bits, Gates, and Programs}
\vspace*{1mm}
\noindent
\textcolor{black}{\textbf{Quantum Bits (Qubits).}\hspace*{1mm} The fundamental unit of quantum computing is the quantum bit or qubit. A qubit can exist in a \emph{ground state} $\ket{0}$, similar to classical 0, or an \emph{excited state} $\ket{1}$, similar to classical 1.}
Unlike classical bits, qubits can exist in a superposition state, representing both $\ket{0}$ and $\ket{1}$ simultaneously. This state is described as $\ket{q} = \alpha\ket{0} + \beta\ket{1}$, where $\alpha$ and $\beta$ are complex numbers that satisfy $|\alpha|^2 + |\beta|^2 = 1$. Here, $|\alpha|^2$ and $|\beta|^2$ denote the probabilities of observing states 0 and 1, respectively.

\begin{figure}[htbp]
    \begin{minipage}[t]{0.5\linewidth}
        \centering
        \includegraphics[width=1\textwidth]{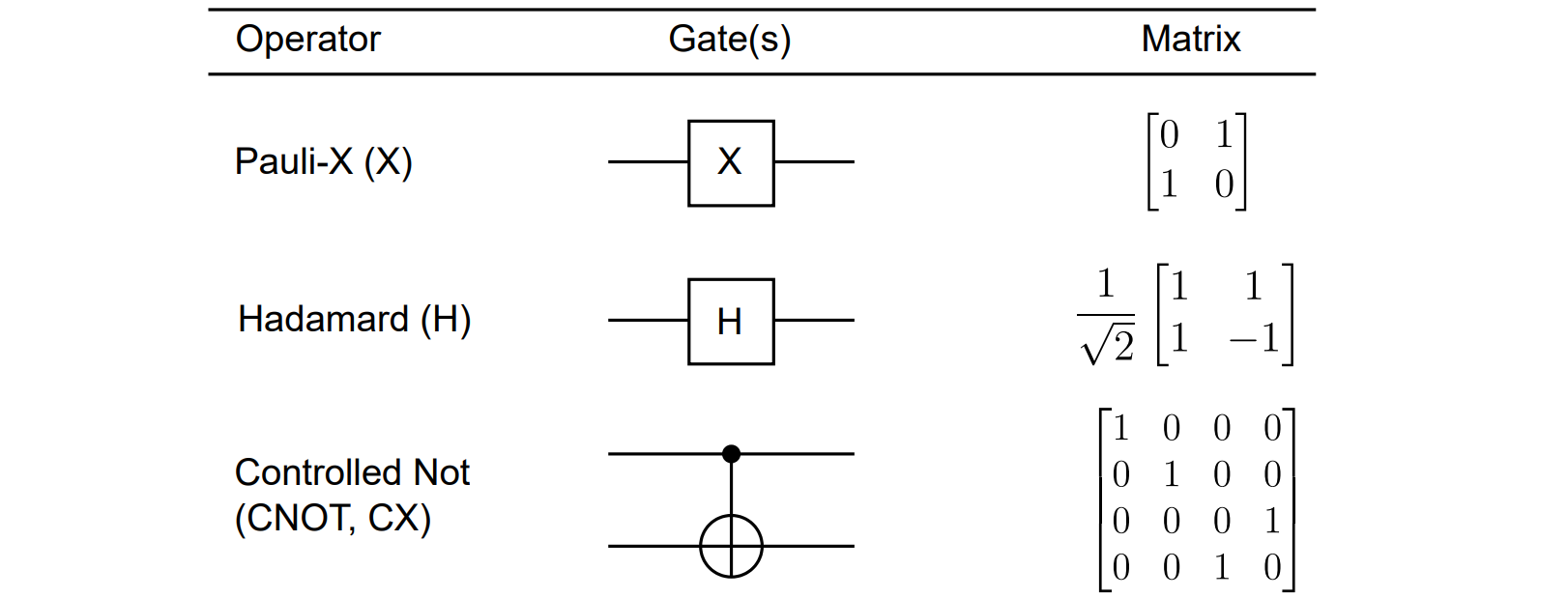}
        \caption{Common quantum gates and matrix.}
        \label{fig:gate}
    \end{minipage}%
    \begin{minipage}[t]{0.5\linewidth}
        \centering
        \includegraphics[width=1\textwidth]{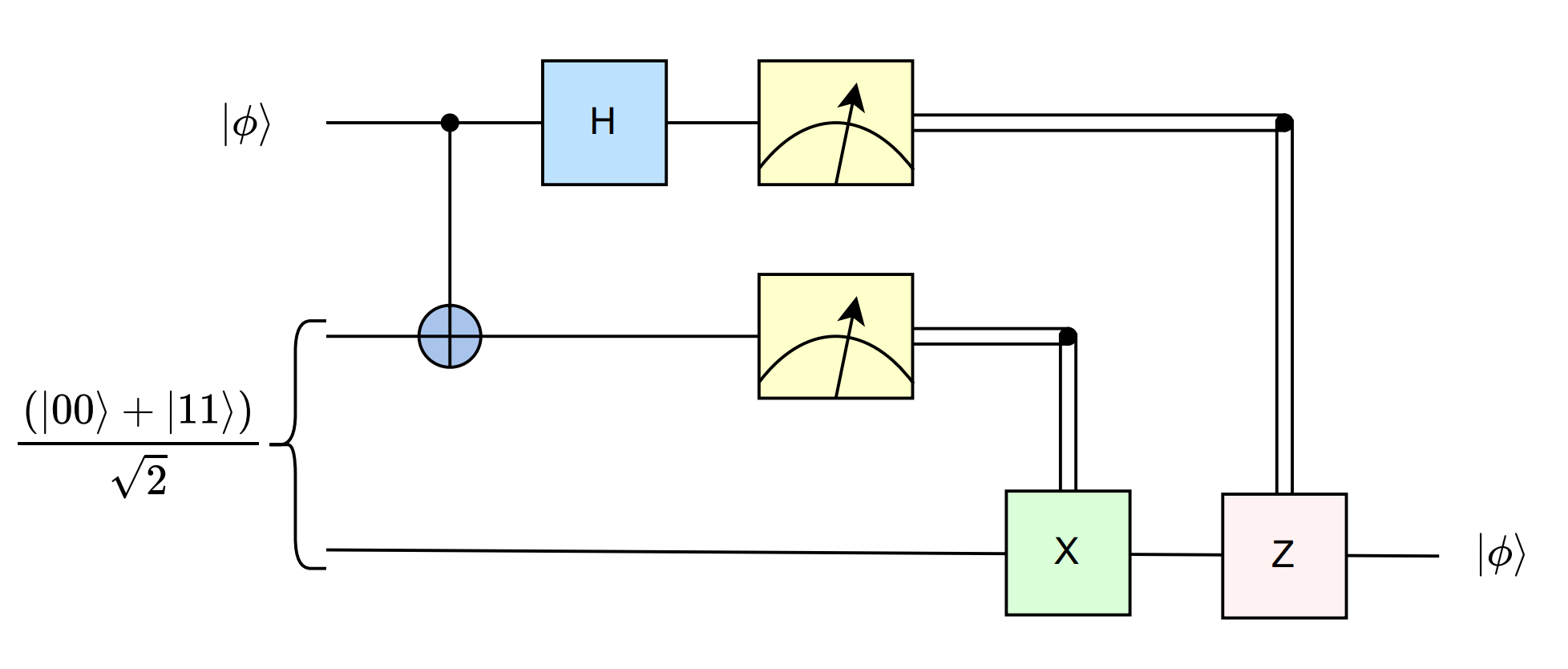}
        \caption{Quantum teleportation.}
        \label{fig:tele}
    \end{minipage}
\end{figure}

\vspace*{1mm}
\noindent
\textcolor{black}{\textbf{Quantum Gates (Operations).}\hspace*{1mm} Quantum gates are essential components in quantum computing, performing operations on qubits such as rotation, inducing superposition and establishing control relationships among qubits. Each gate is defined by its input parameters, which determine the target qubits and the specifics of the operation. Some common quantum gate operations are depicted in \autoref{fig:gate}. }
\textcolor{black}{For example, for a quantum state $\ket{q}=\alpha\ket{0}+\beta\ket{1}$, $X$ gate acting on $\ket{q}$ produces a new quantum state $\ket{q_{new}}=X\ket{q} = \beta\ket{0}+\alpha\ket{1}$. Due to the property of the unitary operator, most quantum operations do not affect the normalization conditions of the quantum state.}

\begin{figure*}[htb]
\centerline{\includegraphics[width=0.6\linewidth]{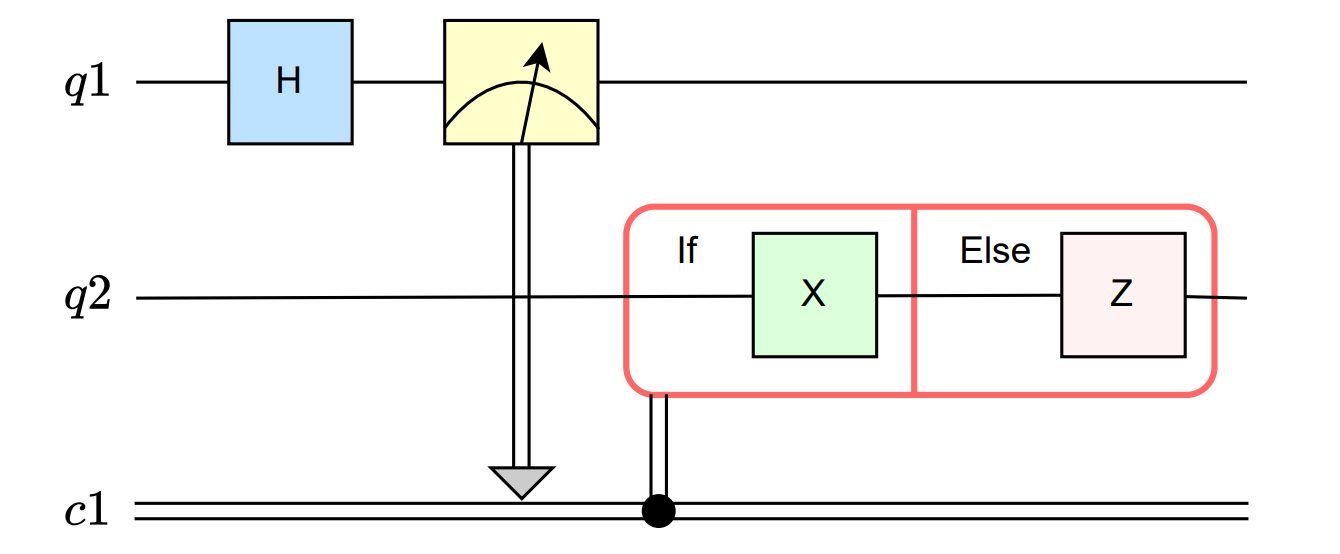}}
\vspace{-0cm}
\caption{An example of a dynamic circuit in Qiskit.}
\label{fig:dynamic}
\end{figure*}

\noindent
\textbf{Quantum Programs.}\hspace*{1mm}
Quantum programs are the building blocks of quantum computation, created by combining qubits and quantum gates. When classical and quantum programs are combined, the classical program statements may involve new quantum logic, such as conditional statements. \textcolor{black}{For example, quantum teleportation shown in \autoref{fig:tele} is an important algorithm in quantum computing that transmits indescribable quantum states from one qubit to another. To achieve this, the program must not only use quantum operations to establish entanglement but also selectively apply the $X$ gate and the $Z$ gate to the final qubit. Whether these two operations are applied depends on the measurement results of the first two qubits. Furthermore, the introduction of dynamic circuits~\cite{IBMQuantumGuide}, shown in \autoref{fig:dynamic}, confirms that such conditional statements controlled by measurement results have become part of modern quantum programs.}
This type of conditional statement is called a quantum control statement in this paper. 

\subsection{Quantum State and Measurement}

\noindent
\textbf{Quantum States.}\hspace*{1mm}
\textcolor{black}{A quantum state encapsulates the information of a quantum system through a vector representation. For a single qubit, its state space is a two-dimensional vector defined as $\ket{q} = \alpha\ket{0} + \beta\ket{1}$. In the context of an \emph{n}-qubit system, with $B = \{\ket{0},\ket{1},\ket{2} \ldots, \ket{2^{n}-1}\}$ representing the \emph{computational basis}, the state of the system is expressed as:}

\vspace{-0cm}
\begin{equation*}
\ket{\psi} = {\textstyle \sum_{\ket{x} \in B}} 
\  \alpha_{x} \ket{x},
\end{equation*}

\vspace{-0cm}\noindent
where $x$ denotes the measurement outcome associated with the basis state $\ket{x}$. Here, each $\alpha_{x}$ is a complex number in $\mathbb{C}$ that describes the magnitude and phase of the quantum states, satisfying the normalization condition $\sum_{\ket{x}\in B}{} |\alpha_{x}|^{2} = 1$.
\textcolor{black}{We can describe the quantum state $\ket{\psi}$ by the vector $v=[\alpha_{0}, \alpha_{1}, \alpha_{2},... \alpha_{2^{n}-1}] \in \mathbb{C}^{2^{n}}$.}

\vspace*{2mm}
\noindent
\textbf{Quantum Measurement.}\hspace*{1mm}
\textcolor{black}{Quantum measurement is a critical process that converts quantum states into classical bit values. In constructing a quantum program, quantum states are manipulated using a series of quantum gates. However, classical computing systems do not directly interpret these states. Thus, the measurement operation $measure()$ is used to translate the quantum state into a classical counterpart. The original quantum state collapses upon measurement, yielding a specific classical outcome based on the state's probability distribution. For example, measuring the Bell state ($\frac{1}{\sqrt{2} } \ket{00} + \frac{1}{\sqrt{2} } \ket{11}$) will yield 00 or 11, each with a 50\% ($|\frac{1}{\sqrt{2}}|^2$) probability. Consequently, the design of an effective quantum program focuses on maximizing the probability of achieving the desired outcome.}

\subsection{Concolic Testing}

Concolic testing~\cite{godefroid2005dart,sen2005cute,majumdar2007hybrid} merges the practicality of concrete execution with the thoroughness of symbolic execution, providing an effective automatic test case generation method. This technique begins with the program running on a specific input to trace an execution path. Concolic testing then heuristically alters the path condition of this execution and employs a constraint solver to process the new constraint. This results in a new concrete input being generated, allowing the exploration of an alternative execution path.
\textcolor{black}{Concolic testing has become a practically usable methodology in software testing, supported by a range of tools~\cite{godefroid2005dart,sen2005cute, majumdar2007hybrid,irlbeck2015deconstructing} designed for various programming languages.}

\subsection{Qiskit}
Qiskit~\cite{qiskit}, a widely used quantum programming language, is an SDK based on Python for constructing and manipulating quantum circuits. It provides features such as quantum circuit creation, quantum state manipulation, and simulation. Quantum manipulations in Qiskit are performed using built-in functions of the quantum circuit object.

\autoref{fig:example} shows an example of the quantum teleportation program written in Qiskit.
\textcolor{black}{To show both statements controlled by classical variables and statements controlled by quantum state measurement results, we made some minor adjustments to the original quantum teleportation program. The function $bob\_process$ accepts the classical variable $alice$ and the quantum variable $qc$ as input and is executed as the main function to construct the quantum circuit. For example, line 15 describes the process of adding an $X$ gate to the qubit ($index=1$) in $qc$. 
Since the execution of the quantum program requires a real computer or simulator, we integrate this part of the code into the function $check\_measure$. For example, on line 14, the function $check\_measure$ performs a single measurement of the qubit ($index=0$) in the quantum variable $qc$ and outputs the measurement result.}

\vspace{2mm}
\begin{figure}[htb]
\footnotesize
\begin{lstlisting}[numbers=left, xleftmargin=2em]
def check_measure(qc, index, register):
    qc.measure(index, register)
    simulator = Aer.get_backend('aer_simulator')
    compiled_circuit = transpile(qc, simulator)
    job = simulator.run(compiled_circuit, shots=1).result().get_counts()
    temp_result = list(job)[0][::-1]
    result = ""
    for i in register:
        result += temp_result[i]
    return result
    
def bob_process(alice, qc):
    # teleportation processing
    if check_measure(qc, [0], [0]) == 1:
        qc.x(1)
    if alice == 1:
        qc.z(1)
        
    # Bob's subsequent operations
    qc.h(1)
    if check_measure(qc, [1], [1]) == 0:
        return 0
    else:
        return 1
\end{lstlisting}
\centering
    \vspace{-0cm}
    \caption{The quantum teleportation program written in Qiskit.}
    \label{fig:example}
\end{figure}

\section{Motivating Example}\label{sec:example}

\textcolor{black}{This section demonstrates the utility of quantum concolic testing through an illustrative example, as depicted in ~\autoref{fig:example}.}

\textcolor{black}{The function \emph{bob\_process} describes the subsequent operations that will be performed on the Bob side in quantum teleportation. The function \emph{check\_measure} describes the fixed process of the measurement for \emph{qc} by calling the simulator in Qiskit.}

\textcolor{black}{In this example, Bob needs to determine if an additional X gate operation (line 15) is required based on the measurement (line 14) of the qubit (index 0) in the current circuit \emph{qc}, and then determine if an additional Z gate operation (line 17) is required based on the parameter \emph{alice} transmitted from Alice. After this, the qubit (index 1) will become the one Alice wants to transmit. The following lines, 20-24, are the subsequent operations performed by Bob.}

\textcolor{black}{In this example program, we notice multiple execution paths, all of which are potentially buggy. Therefore, we aim to obtain a series of test cases (\emph{alice},~\emph{qc}) to detect the presence of bugs in each execution path.}
By inspecting the \textsf{if} statement, we realize that the generation of multiple execution paths arises from two kinds of conditional statements: the classical conditional statement (line 16) and the quantum control statement (lines 14 and 21), as it is called here.

\textcolor{black}{For the classical conditional statement (line 16), we consider classical concolic testing, which can effectively generate $alice = 1$ in classical programs facing the same statement. However, classical concolic testing faces quantum hurdles in this context. \emph{qc} as an unknown symbol type can interfere with the execution process of concolic testing. Hence, we design a new symbolic object \emph{sqc} for quantum variables that can identify and record all operations on quantum variables in a program. This enables classical and quantum variables to perform computations without affecting each other.}

\textcolor{black}{For the quantum control statements (lines 14 and 21), we design constraint generation methods for quantum control statements. Since the semantics of \emph{measure()} involve obtaining the value of specified qubits, we fix the specified qubits to satisfy the condition and obtain the complete computational basis using the tensor product. We generate constraints that ensure other computational bases in the quantum state output with probability 0 to address the ambiguity. We then express the quantum state in terms of a sequence of classical variables and generate the final sequence using all the quantum operations in the execution path. By computing the final sequence and the constraints, we can obtain the quantum state that triggers the quantum control statement.}

\textcolor{black}{So, for the function \emph{bob\_process} (referred to as $B$) in the example, our method initially assumes ($alice = 0$, $qc = \ket{00}$) as the first test case and creates symbolic objects ($alice_0$, $sqc$) for the input variables ($alice$, $qc$).
} 
During this execution, we obtain a result \emph{0} and a path constraint $alice_0 \neq 1 \land sqc[][0] \notin ["1"] \land sqc[h(1)][1] \in ["0"]$ for this result, guiding us to explore another branch. Using these constraints, our method generates a new condition $alice_0 \neq 1 \land sqc[][0] \notin ["1"] \land sqc[h(1)][1] \notin ["0"]$ and submits it to a quantum constraint solver. 
\textcolor{black}{The solution, ($a=0$, $qc = (0.39-0.44i)\ket{00} + (0.52-0.62i)\ket{01}$), is recorded as a new test case.} 
Running $B$ with this new test case subsequently exposes a new execution path for a new result \emph{1}.
Some of the details about this example are presented in the Appendix \ref{smt}.

\textcolor{black}{In this process, we also find that the probabilistic output of the quantum program results in the extremely unstable triggering of each execution (line 21). The execution paths that each test sample can trigger are random. This situation arises because the \emph{measure()} statement can only obtain a single value of the quantum state and cannot characterize the probabilistic properties of the quantum state.
This makes Bob unable to make more precise requirements for the quantum states that Alice transmits. To solve this problem, we provide three additional methods for constraint generation to characterize the probabilistic nature of quantum states, described in detail in Section \ref{quantum_condtion}.}

\section{Methodology}\label{sec:approach}
We next present our methodology for quantum concolic testing.

\subsection{Workflow}

\begin{figure*}[htb]
\centerline{\includegraphics[width=1\linewidth, height = 0.4\textwidth]{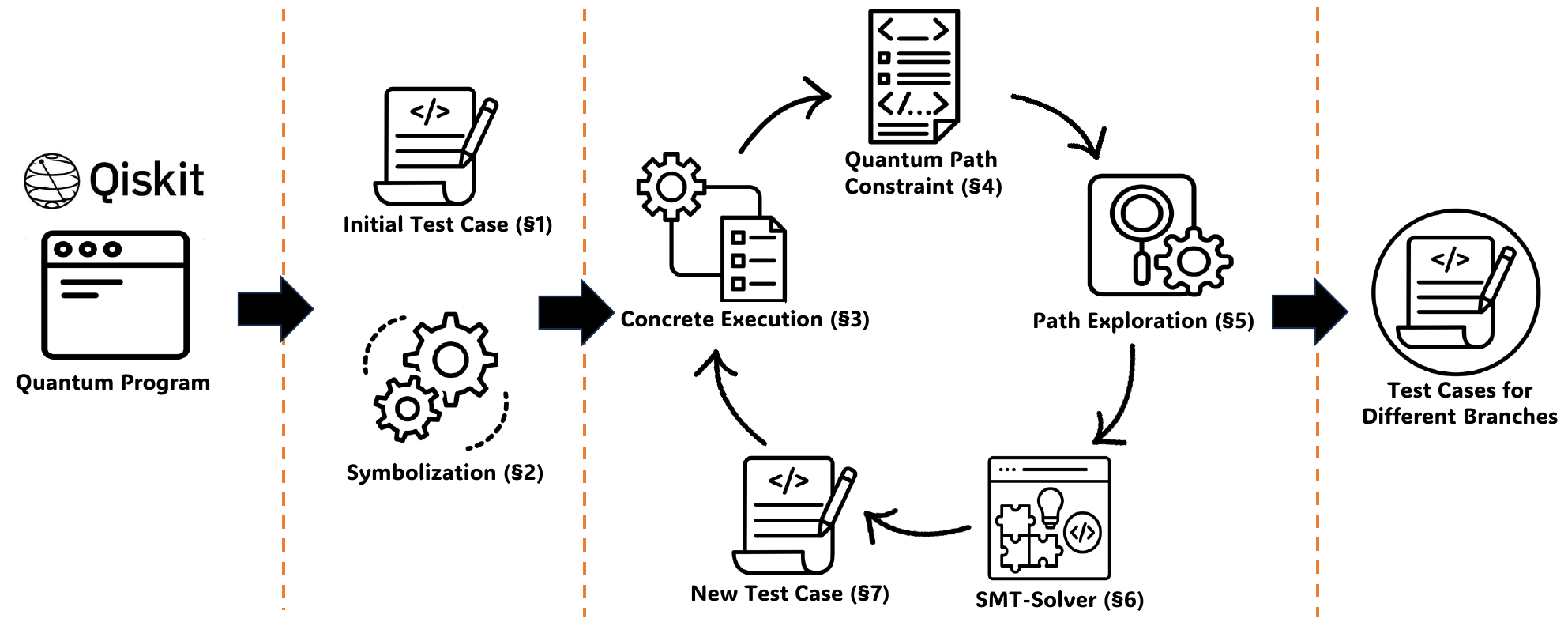}}
\vspace{-0cm}
\caption{Workflow of quantum concolic testing framework.}
\label{fig:overview}
\end{figure*}

Our quantum concolic testing framework is designed to uncover input samples that effectively trigger all branches within a quantum program. In the following, we detail our strategy for generating a comprehensive test suite tailored for quantum programs, as illustrated in \autoref{fig:overview}.

Our framework consists of seven components: 1) For the first concrete execution, we generate the initial test case based on the classical and quantum input parameters of the target quantum program. 2) We create the symbolic objects corresponding to the type of input parameters. 3) We perform a concrete execution to obtain the current execution path and results. 4) Based on the current execution path, we utilize the devised quantum constraint generation methods to produce the current quantum path constraint. 5) To explore more execution behaviors, we modify the current path constraint and generate a new one. 6) We use an SMT solver to solve the new path constraint and generate a new test case based on the solution. 7) We execute a concrete execution using the new test case to detect whether a new branch is triggered.

The following sections will illustrate quantum constraint generation methods, symbolic objects for quantum variables, and quantum path exploration methods.


\subsection{Quantum Condition}
\label{quantum_condtion}
Current quantum control statements base their judgments on a single observation of a quantum circuit's behavior.
\textcolor{black}{For the generic measurement statement \emph{measure()} in quantum languages, let $Q_m$ be the set of target observation qubits, and $O_r$ be the set of expected observations. Thus, quantum control statements in quantum programs can be written as $measure(Q_m) == O_r$. For an n-qubit system, we assume that $N$ is the set of qubit indices, $S$ is the set of all potential observations, and $\ket{\phi}$ is the quantum state when the quantum program executes the quantum control statement. Hence, we will produce the following constraint formula:}

\vspace*{1mm}
\begin{large}
\begin{center}
$\textstyle \bigwedge_{a \in S\setminus S_r} (\left \langle a  | \phi  \right \rangle =0) = True$
\end{center}
\end{large}
\vspace*{1mm}

\noindent
\textcolor{black}{where $S_r = \sum_{i \in O_r} i\otimes I_{N\setminus Q_m}$ is the set of n-qubit states that satisfy the observation condition.}

When we solved the branches generated by the statement \emph{measure()}, we realized the drawbacks of the randomness of the measurement.
The inherently probabilistic nature of quantum program outputs makes the direction of branching within these statements uncertain. This uncertainty manifests itself in several significant issues: Firstly, it is challenging to determine whether the behavior of a quantum state aligns with predetermined requirements using existing quantum control statements. Even a random quantum state might, by chance, trigger a branch, misleading developers into thinking the state conforms to the intended program design. This misinterpretation can lead to persistent errors, which are especially problematic when scaled up.
For instance, if the expected output quantum state of the program is a Bell state, the correctness of the program is difficult to determine solely by observing whether the outcome is 00 or 11.
%
Secondly, the inability of \emph{measure()} to support developers in specifying the requirements of quantum states turns each quantum program design endeavor into a gamble. This lack of quantifiable metrics significantly hampers the efficiency of quantum program development, as it prevents a clear understanding and assessment of the target quantum state's properties.



To avoid the uncertainty of the "blind box" program, we impose stricter constraints on the quantum states. Subsequently, we design three constraint generation methods based on the probabilistic outputs of quantum states, as described below.


\begin{itemize}[leftmargin=1.5em]
\setlength{\itemsep}{3pt}

\item \textbf{Check\_state\_eq (equal to)}: 
Check whether the overall distribution of the quantum state under measurement satisfies the expected probability distribution requirements. For example, we expect the distribution $D$ of the quantum circuit $qc$ to satisfy an acceptable error margin $\delta$.
This conditional statement can be written as $check\_state\_eq(qc, D, \delta)$ and generates the constraint formula as follows:

\vspace*{1mm}
\begin{large}
\begin{center}
${\textstyle \bigwedge_{a \in S} (\left |   |\left \langle a  | \phi  \right \rangle|^2 - D_a\right | < \delta )} =True $
\end{center}
\end{large}
\vspace*{1mm}

\textcolor{black}{where $\ket{\phi}$ is the quantum state of $qc$, $S$ is the set of all  observations and $D_a \in D$ is the expected probability of the observation $a$.}

\item \textbf{Check\_state\_gt (larger than)}: 
Set the lower limit on the probability of specific observations of the quantum state under measurement.
\textcolor{black}{For instance, we expect the quantum variable $qc$ to have a higher probability of outputting $a$ than $p$ with an acceptable error $\delta$, written as an expected state pair $(a,p)$ and $P$ is the set of all expected state pairs.
This conditional statement can be written as $check\_state\_gt(qc, P, \delta)$ and generates the constraint formula as follow:}

\vspace*{1mm}
\begin{large}
\begin{center}
${\textstyle \bigwedge_{s \in P \wedge  s=(a_s, p_s)} (  |\left \langle a_s  | \phi  \right \rangle|^2  > p_s -\delta )} =True $
\end{center}
\end{large}
\vspace*{1mm}

\textcolor{black}{where $\ket{\phi}$ is the quantum state of $qc$.}

\item \textbf{Check\_state\_lt (less than)}: 
Set the upper limit on the probability of specific observations of the quantum state under measurement.
\textcolor{black}{For instance, we expect the quantum variable $qc$ to output $a$ with a probability less than $p$ under an acceptable error $\delta$, written as an expected state pair $(a,p)$ and $P$ is the set of all expected state pairs.
This conditional statement can be written as $check\_state\_lt(qc, P, \delta)$ and generates the constraint formula as follow:}

\vspace*{1mm}
\begin{large}
\begin{center}
${\textstyle \bigwedge_{s \in P \wedge  s=(a_s, p_s)} (  |\left \langle a_s  | \phi  \right \rangle|^2  \le p_s +\delta )} =True $
\end{center}
\end{large}
\vspace*{1mm}

\textcolor{black}{where $\ket{\phi}$ is the quantum state of $qc$.}

\end{itemize}

\begin{figure*}[htb]
 \centerline{\includegraphics[width=0.95\linewidth, height = 0.25\textwidth]{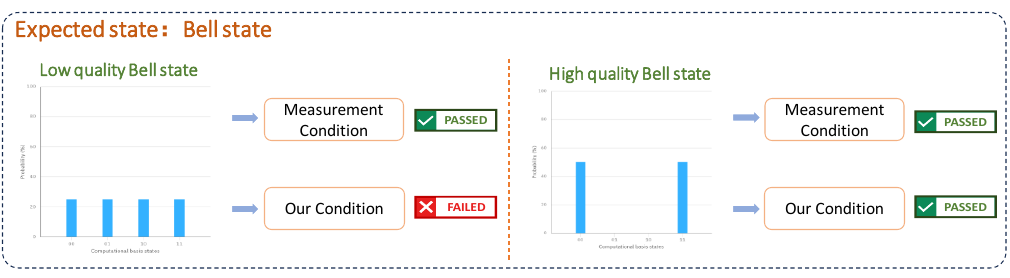}}
\caption{For the two quantum states with different output distributions, the different results between the measurement condition and our condition.}
\label{fig:low_high}
\end{figure*}

We evaluated the quality of a quantum state based on the conditions we have devised. A high-quality quantum state has a distribution that closely matches the expected required distribution. This means that the quantum state can produce the expected result with a very high probability, even with a small number of observations. On the other hand, a low-quality quantum state significantly deviates from the expected requirements. It requires many quantum resources to produce the expected result. By applying these three conditional forms, we aim to eliminate a significant portion of low-quality quantum states that do not meet the expectations of program design, as illustrated in \autoref{fig:low_high}. 

\subsection{Quantum Symbolic Object}
\label{symbolic_object}
In the classical concolic testing~\cite{godefroid2005dart, sen2005cute, majumdar2007hybrid}, symbolic variables are generated for the target variables to construct constraints on execution paths. The symbolic variables record all arithmetic operations on the variables during program execution and apply these operations to generate constraints. For example, if the program statement is $x=x+1$ and the condition is $x==5$, the final constraint generated is $x+1==5$. However, the logic of operations in a quantum program differs from the arithmetic logic of classical variables. \textcolor{black}{The operations in quantum computation act on the qubits in the quantum circuit, e.g., $qc.h(0)$ indicates that a Hadamard gate is appended to the first qubit.
This operational logic of quantum computation does not appear in classical computation, and quantum variables cannot be input directly into the constraint solver. Therefore, the operations of quantum variables in the execution path cannot be analyzed using the symbolic approach of classical concolic testing.}

\textcolor{black}{To solve this problem, we develop a new symbolic object for quantum variables in quantum programs. Quantum program operations usually involve the variable itself and parameterize the indices of the target qubits. Therefore, we opt to treat the entire circuit as a quantum variable. All operations performed on the quantum circuit during the current execution path are recognized and recorded in $operation\_list$, a built-in attribute of the quantum symbolic object. This list varies across different execution paths, capturing different quantum operations for each.}
\textcolor{black}{Inspired by SymQV~\cite{bauer2023symqv}, we express the quantum state $\ket{\phi} = \sum_{\ket{x}\in B}{} \alpha_{x} \ket{x}$ in terms of the sequence $(a_0, ..a_n, b_0,..b_n)$ generated by $\alpha_{x}=a_x+b_{x}j$ corresponding to each computational basis, which $a_i$ and $b_i$ are the real numbers. Based on the number of qubits in the program, we set the initial sequence $(a^{0}_0, ..a^{0}_n, b^{0}_0,..b^{0}_n)$ 
With a concrete execution, each quantum operation generates a new sequence $(a_{0}^{k},..a_{n}^{k}, b_{0}^{k},..b_{n}^{k})$, which $a_{i}^{k}$ and $b_{i}^{k}$ are generated by a linear combination of $a_{j}^{k-1}$ and $b_{j}^{k-1}$ ($i,j \in {0..n}$). Similar to SSA, these combinations will all exist as part of the SMT file. When the final execution reaches the quantum control statement, we use formulas to generate the constraint expression. All expressions from this entire process will be used in the constraint solver.}
\textcolor{black}{Under this processing flow, the transformation logic for all quantum operations is preserved in the generated SMT file.}

\textcolor{black}{In addition, since most quantum operations have matrix representations, we can construct a matrix $U_{all}$ that encodes all quantum operations along the execution path by integrating the operations recorded in the $operation\_list$. We then apply $U_{all}$ to transform the initial sequence $(a^{0}_0, ..a^{0}_n, b^{0}_0,..b^{0}_n)$ into a new sequence $(a^{f}_0, ..a^{f}_n, b^{f}_0,..b^{f}_n)$. This transformation process is included in the SMT file to replace the constraint expressions previously generated for each quantum operation. Through this process, for an execution path of an (n-qubit, m-operation) program, the number of variables in the SMT file is reduced from $(m+1)\times2^{n}$ to $2^{n+1}$, thus improving the efficiency of constraint solving. However, it is important to note that while this simplification reduces the number of variables, it also increases the complexity of individual constraints and makes it harder to trace the relationship between constraints and specific quantum operations.}

Moreover, we built the three quantum control statements defined in Section \ref{quantum_condtion} into the quantum symbolic objects we designed.

\subsection{Quantum Constraint}
\label{quantum constraint}
\textcolor{black}{In this section, we will introduce the generation and exploration of quantum constraints based on the quantum conditions outlined in Section \ref{quantum_condtion} and the quantum symbolic objects described in Section \ref{symbolic_object}.}

\vspace*{2mm}
\noindent
\textbf{Generation.}\hspace*{1mm}
\textcolor{black}{To generate complete quantum constraints, we first generate preset constraints based on the quantum symbolic object that has recorded the operations in the execution path. The corresponding constraint formulas are then generated according to the quantum condition faced, respectively, and documented as follows:}
%

\vspace*{1mm}
\begin{center}
$measure\xrightarrow{} (qc[operations] \in [observations_{expected}])$\\
$check\_state\_eq \xrightarrow{} (qc[operations] = [probability_{all}])$\\
$check\_state\_gt \xrightarrow{} (qc[operations] > [probability_{par}])$ \\
$check\_state\_lt \xrightarrow{} (qc[operations] \leq [probability_{par}])$
\end{center}
\vspace*{1mm}

\noindent
\textcolor{black}{where $qc$ is the name of the quantum symbolic variable and $operations$ is the quantum operations on the execution path read from the built-in attribute $operation\_list$. $probability_{all}$ and $probability_{par}$ are the complete output distribution of the entire quantum state and a list consisting of specific output outcomes and the corresponding probabilities, respectively.}

\vspace*{2mm}
\noindent
\textbf{Exploration.}\hspace*{1mm} 
\textcolor{black}{For each conditional statement, we have generated a constraint $Q$, which corresponds to the branch choice.
At this point, $\neg Q$ represents the constraint to branch in the other direction.
Therefore, our exploration method will transform the path constraint $P\land Q$ of the current execution path into $P \land \neg Q$ to explore another execution path. 
For the quantum conditions that we designed, we document the mutual conversion of $\in$ and $\notin$ for $measure$, $=$ and $\neq$ for $check\_state\_eq$, and the mutual conversion of $>$ and $\leq$ for the other quantum control statements.}

\textcolor{black}{To avoid exploring the same execution paths repeatedly, we build a forked tree for the target program. Each conditional statement is recorded as a non-leaf node and is divided into two sub-branches based on the execution result. When both sub-branches have been explored, we mark the non-leaf node as totally explored. We generate a path from the terminal leaf node to the root node based on this forked tree, and all nodes in this path are not marked as completely explored. All constraints in the non-leaf nodes in this path are then concatenated with $\land$ to generate the constraint formulas for the target explored branch.}

\textcolor{black}{We pass the newly acquired constraints directly to the quantum constraint solver, which generates the corresponding SMT expressions and appends them to the SMT file produced from the quantum state and quantum operations along the execution path. Due to the presence of rational numbers and quadratic equations, we use the \emph{dReal} solver~\cite{10.1007/978-3-642-31365-3_23} to perform SMT solving. We then process the computational results of \emph{dReal} to generate new quantum test cases. It is important to note that, due to the probabilistic nature of quantum program outputs, the test cases generated by \emph{dReal} may fail to trigger the expected new branch within a small number of executions. In such cases, solving the same SMT file multiple times will produce identical test cases, making it inefficient to explore new branches. To address this, we convert ineffective test cases generated from the current SMT file into additional constraints and append them to the SMT file. For example, if \emph{dReal} computes that $a_{i}^{0} \in [0.5, 0.6]$, but the corresponding test case fails to trigger the target branch, we add a new constraint $a_{i}^{0} \notin [0.5, 0.6]$ to the SMT file. This ensures that the SMT file changes each time the same new branch is explored, allowing the generated test cases to be different.}


\subsection{Algorithm}
\textcolor{black}{This section explains how our framework uses quantum concolic testing to generate input samples for quantum programs.}
\textcolor{black}{Algorithm \autoref{alg1} presents the overall flow of the process.}

\begin{algorithm}
	\renewcommand{\algorithmicrequire}{\textbf{Input:}}
	\renewcommand{\algorithmicensure}{\textbf{Output:}}
	\caption{: Quantum Concolic Testing}
	\label{alg1}
        \begin{footnotesize}
	\begin{algorithmic}[1]
        \REQUIRE $P\gets$ target quantum program \\
                \ \ \ \ $S_{results}\gets$ set of expected results \\
                \ \ \ \ $i_{max}\gets$ number of maximum iterations \\
                \ \ \ \ $r\gets$ number of repetitive executions
        \ENSURE $S_{inputs}\gets$ set of test cases
		\STATE $S_{inputs},\  inputs\_list,\  results = [], [], []$
            \STATE $input_{init} = create\_initial(P)$ \ \ \textcolor{teal}{//Initialize an input for P}
		\STATE $i = 0$ \ \ \ \  \textcolor{teal}{// Number of iterations}
            \STATE $inputs\_list.append(input_{init})$
            \STATE $symbol = symbolization(P)$ \\
		\REPEAT
            \STATE $input = inputs\_list.pop()$ \\
            \STATE $symbol.update(input)$ \\
            \FOR{$j\gets 0$ to $r$}
            \STATE $exe\_result = concrete\_execution(P, symbol)$ \\
             \textcolor{teal}{// Generating path constraints using symbolic variables}
            \STATE $path\_constraint = generate\_constraint(P, symbol)$\\
             \textcolor{teal}{// Detect if a new execution branch is triggered}
            \IF{$exe\_result$ not in $results$}
            \STATE $results.append(exe\_results)$
            \STATE $S_{inputs}.append(input)$
            \STATE break
            \ENDIF
            \ENDFOR \\
            \textcolor{teal}{// Generating a new path constraint expression}
            \STATE $new\_constraint = exploration(path\_constraint)$ \\
             \textcolor{teal}{// Using the solver to find the input sample}
            \STATE $input_{new} = quantum\_constraint\_solver(new\_constraint)$
            \STATE $inputs\_list.append(input_{new})$
            \STATE $i = i + 1$
		\UNTIL $explore\_complete(results, S_{results})$ or $i > i_{max}$
	\end{algorithmic}  
        \end{footnotesize}
\end{algorithm}

\textcolor{black}{For the quantum program given for the analysis, we will first create an initial test case as the start of the first concrete execution (line 2). For classical and quantum variables, we initialize them to 0 and $\ket{00..0}$, respectively. We will symbolize all input variables to generate the corresponding symbolic objects (line 5). Then we go inside concolic testing and update the values for concrete executions in symbolic objects (line 8). Since quantum programs are probabilistic outputs, we execute concrete executions $r$ times (lines 10) for a test case to fully explore all potential outputs and check whether a new branch is triggered (lines 12-14). For each execution, we generate a path constraint on the current execution path (line 11).}
\textcolor{black}{Based on the current path constraint, we utilize the path exploration method described in Section~\ref{quantum constraint} to yield a new constraint for a new branch (line 18). Then, this constraint is solved by the SMT-solver, and we generate a new test case based on this solution (line 19). This new test case is then used in the next path exploration (line 20). The whole process will end after all branches have been explored or the maximum number of searches has been reached.}


\section{Experimental Evaluation}\label{sec:eval}
To evaluate the effectiveness of our proposed algorithm, we implement our quantum concolic testing framework in Python based on the Qiskit Library~\cite{qiskit}. In the experiments, we intend to answer the following questions:

\definecolor{bleudefrance}{rgb}{0.19, 0.55, 0.91}

\begin{itemize}[leftmargin=2em]
\setlength{\itemsep}{3pt}

\item \textbf{RQ1}: (Branch Coverage) How effective is quantum concolic testing at improving branch coverage in quantum programs?

\item \textbf{RQ2}: (Efficiency) How efficient is quantum concolic testing at exploring program branches?

\item \textbf{RQ3}: (Quality Performance) What is the quality of the test samples generated by quantum concolic testing?

\item \textbf{RQ4}: \textcolor{black}{(Bug Finding) How effective are test cases generated by quantum concolic testing at finding bugs in quantum programs?}
\end{itemize}

In \textbf{RQ1}, we aim to explore our framework's coverage capability for varying-size quantum programs in terms of branch coverage. 
In \textbf{RQ2}, we will discuss the efficiency of our framework in improving the coverage of the branches compared to other methods within the same time frame.
In Section \ref{quantum_condtion}, we discuss the quality of quantum input samples. Thus, in \textbf{RQ3}, we will quantitatively analyze the performance of the input samples generated by our framework in terms of quality. \textcolor{black}{In \textbf{RQ4}, we aim to evaluate whether the generated test cases can effectively detect faults in quantum programs.}

For the baseline, we compare our framework (\textbf{Concolic}) with three quantum program test case generation approaches, which are random quantum state vector generator (\textbf{Vector}), random quantum circuit generator (\textbf{Circuit}), 
\textcolor{black}{Quito generator (\textbf{Quito}) ~\cite{quito}}
and the QuraTest generator (\textbf{Quratest})~\cite{ye2023quratest}. 
\textcolor{black}{Since many approaches to quantum program testing still do not present a complete framework at this stage, we choose these four test case generation methods for comparison.}
\textcolor{black}{We migrate these four generators into our concolic testing framework by replacing the parts \S 4$ \sim $\S 7 in the workflow to mitigate interference from classical variables.} 
\textcolor{black}{The baseline test generators were originally designed to generate quantum states independently of classical variables. However, the benchmark programs used in our experiments include both classical and quantum variables. By porting the other four methods into our concolic testing framework, we can handle the interplay between classical and quantum variables more effectively. This integration allows test generators to produce inputs that account for both types of variables, leading to a more accurate assessment of their performance.}

\textcolor{black}{As for benchmarks, quantum programs are still in the development stage, and currently, no established benchmark adequately covers the most common quantum operations. Moreover, some existing quantum algorithms \cite{10.1145/237814.237866, Coppersmith2002AnAF,Kitaev1995QuantumMA} focus solely on the construction of quantum circuits without describing the subsequent processing and analysis of the measurement results. This leads to discontinuous interactions between quantum programs and classical programs.}
\textcolor{black}{Hence, We designed a dataset for our experiments comprising Qiskit programs of 32 types of quantum gates, varying program scales (\textbf{S}mall, \textbf{M}edium, \textbf{L}arge) and qubit quantities (1, 2, 3, 4), segmented into 12 sets of experiments. Each experimental set includes 40 Qiskit programs and four hybrid program structures(simple if-statement, nested conditional statements, multiway branch, and multi-parameter control statements). \textcolor{black}{Each program has multiple execution paths corresponding to various quantum circuit structures. The number of quantum operations in each execution path is approximate: Small (S): 5, Medium (M): 10, Large (L): 20.}} 
The acceptable error $\delta $ is randomly set to 0.01 or 0.005. The number of repetitive executions $r$ is 10.

\textcolor{black}{Currently, there is no dedicated quantum constraint solver available. We designed a solver using the framework of an existing quantum symbolic execution tool~\cite{bauer2023symqv}. This solver is not perfect, and as the number of qubits increases, the constraint-solving time increases significantly due to the exponential growth in the number of variables. This results in a substantial increase in the execution time for large-scale samples. Therefore, for this work, we chose to limit the number of qubits to four to verify the feasibility of quantum concolic testing.}
\textcolor{black}{The generated SMT files in \textbf{Concolic} are simplified files after integrating quantum operations. The SMT solving process uses the solver \emph{dReal}. The upper limit of the single SMT solving time of \emph{dReal} is 10,000 s. The $\delta_{sat}$ in \emph{dReal} is 0.05.}

All experiments were run on a computer with Ubuntu 22.04 LTS and Intel Xeon E5-2620v4, 32GB memory, and a 2TB HDD disk. Due to the page limit, the rest of this section mainly discusses the summarized results of our experiment.

\subsection{Branch Coverage}

We evaluated the branch coverage of the test cases generated by our technique. We conducted experiments for the five methods in 12 sets, each containing 40 Qiskit programs. Due to the probabilistic nature of the quantum program output, we set each generated sample to undergo ten repetitions of execution to capture all potential branches triggered. For \textbf{Concolic}, we capped the maximum generated sample count to 50.
\textcolor{black}{For \textbf{Quito}, we utilize all test cases that can be generated.}
For the other three methods, our goal was to obtain 10,000 samples. Subsequently, we calculated the branch coverage achieved by each method. The experimental results are in \autoref{tb:branch coverage}.


\begin{table}[htb]
\footnotesize
\centering
\setlength{\tabcolsep}{5pt}
\renewcommand{\arraystretch}{1}
\caption{Branch coverage results for different methods and program sizes.}
\label{tb:branch coverage}
\begin{tabular}{l|ccc|ccc|ccc|ccc}
\hline
          & 1\_S& 1\_M  & 1\_L & 2\_S& 2\_M  & 2\_L & 3\_S& 3\_M  & 3\_L &
          4\_S& 4\_M  & 4\_L
          \\ \hline
Circuit &  84.58 & 88.47 & 83.61 & 77.71 & 78.54 & 76.67 & 72.78 & 72.95 & 71.97 & 68.33 & 67.81 & 68.02\\
Vector &   87.74 & 91.38 & 86.46 & 76.59 & 79.27 & 77.64 & 68.33 & 68.38 & 69.25 & 66.25 & 65.63 & 66.67\\
Quito &  70.83 & 68.65 & 72.19 & 74.06 & 68.85 & 76.56 & 69.90 & 69.06 & 64.48 & 63.65 & 67.08 & 64.58 \\
Quratest & 84.48 & 88.16 & 82.41 & 78.33 & 80.21 & 78.02 & 77.61 & 76.77 & 72.33 & 71.35 & 70.31 & 69.38\\
Concolic & \textbf{100.0}& \textbf{100.0} & \textbf{99.38}& \textbf{94.48}& \textbf{89.90} & \textbf{85.73}& \textbf{93.65} & \textbf{89.69}& \textbf{83.33} & \textbf{86.98} & \textbf{82.50}& \textbf{74.27}\\
\end{tabular}
\end{table}

\textcolor{black}{The experimental results indicate that \textbf{Concolic} achieved the highest branch coverage rate in all experimental sets. The test case generation method based on quantum path constraints effectively explores various program branches using fewer test cases. In addition, as the number of qubits grows, it leads to an exponential expansion of the input space for quantum variables and causes the other four methods to show a decrease in branch coverage. For \textbf{Concolic}, increasing of qubits number leads to the number of elements in the sequence $(a_{0}^{i},..a_{n}^{i}, b_{0}^{i},..b_{n}^{i})$ which describes the quantum state will also increase exponentially. This computational pressure on the SMT solver results in a decrease in the correctness.
Moreover, the increase in the number of quantum operations has no impact on the test cases generated by the other four methods and, therefore, does not affect their branch coverage. }
\textcolor{black}{Although we have reduced the number of variables in the constraints by integrating the quantum operations on the execution path, an increase in the number of quantum operations causes the integrated matrix to become more complex. \textcolor{black}{This results in constraint expressions involving more variables, thus making the SMT file more complex to solve.}}

\begin{figure*}[htb]
\centerline{\includegraphics[width=0.8\linewidth]{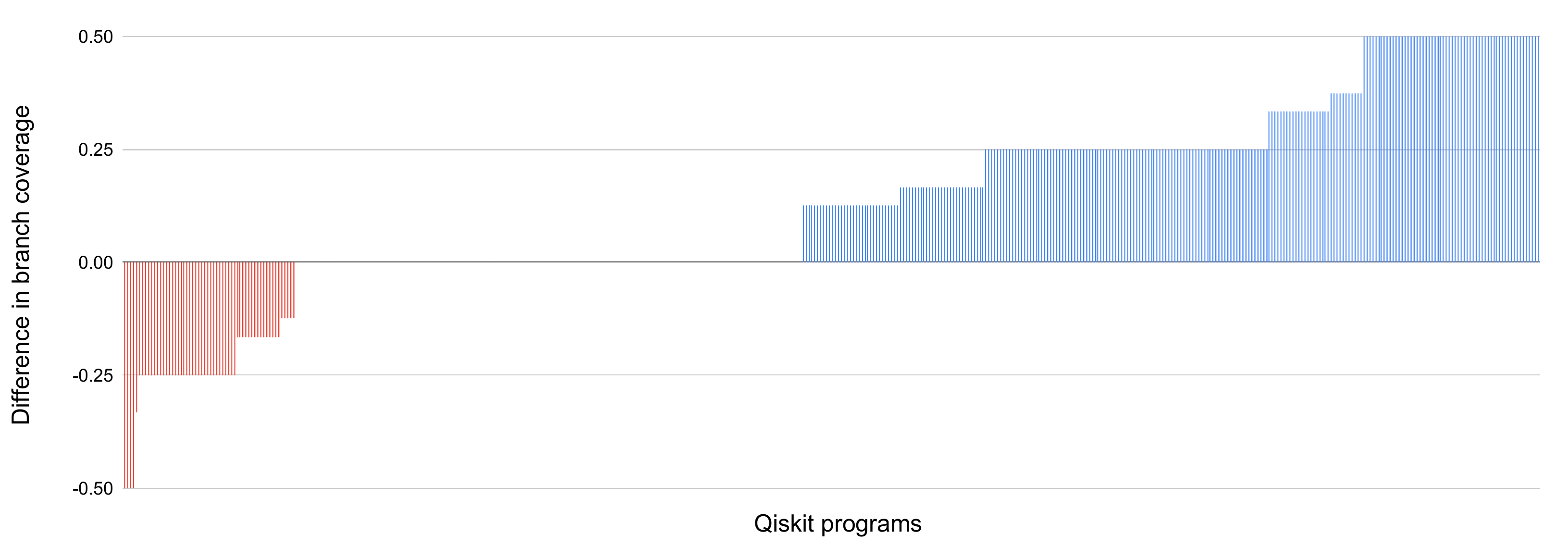}}
\vspace{-0cm}
\caption{Difference in the branch coverage obtained by Concolic and the other three methods.}
\label{fig:difference}
\vspace{-0cm}
\end{figure*}

\textcolor{black}{Additionally, \autoref{fig:difference} shows the difference in branch coverage between Concolic and the best branch coverage of the other four methods for each program in the benchmark (480 programs). The $x$-axis represents the tested Qiskit programs, and the $y$-axis indicates how much branch coverage Concolic improved over the best performance of the other four methods. The left red programs are almost from 4\_Large by \textbf{Quratest}. The blue programs on the right are almost from the 2-qubit and 3-qubit programs.}
\textcolor{black}{For the leftmost red region (-0.50), we analyzed the experimental results. Since we set the single SMT solving time of \emph{dReal} solver to 10000s, for the constraints that exceed this time limit, our framework will abandon further exploration of this branch. This situation can be effectively mitigated by raising the upper bound of the single SMT solving time.}


\begin{tcolorbox}[size=title,rightrule=1mm, leftrule=1mm, toprule=0mm, bottomrule=0mm, arc=0pt,colback=gray!5,colframe=bleudefrance!75!black,breakable]
{ \textbf{Answer to RQ1: } 
In terms of branch coverage, 
\textbf{Concolic} achieves superior branch coverage compared to other methods. We also found that the performance of \textbf{Concolic} is affected by the size of the program, while the other methods are almost unaffected. All generation methods are affected by changes in the number of qubits.
}
\end{tcolorbox}

\subsection{Efficiency}
\begin{figure*}[htb]
    \centering
    \vspace{-0.4cm}
    \subfloat[1\_Small]{\includegraphics[width=0.33\hsize ]{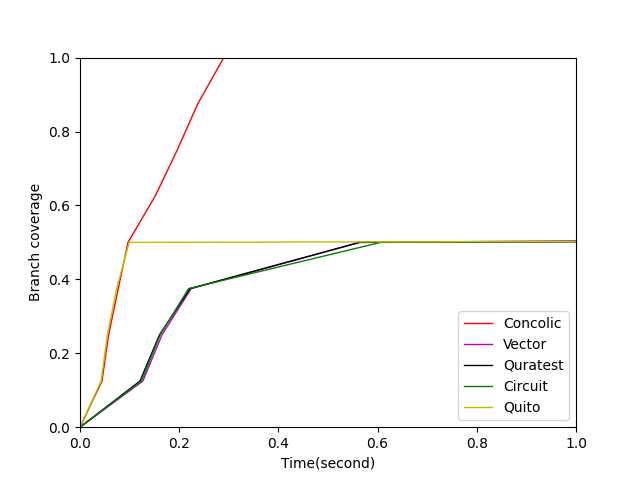}\label{fig: 1_S}}  
    \subfloat[1\_Medium]{\includegraphics[width=0.33\hsize ]{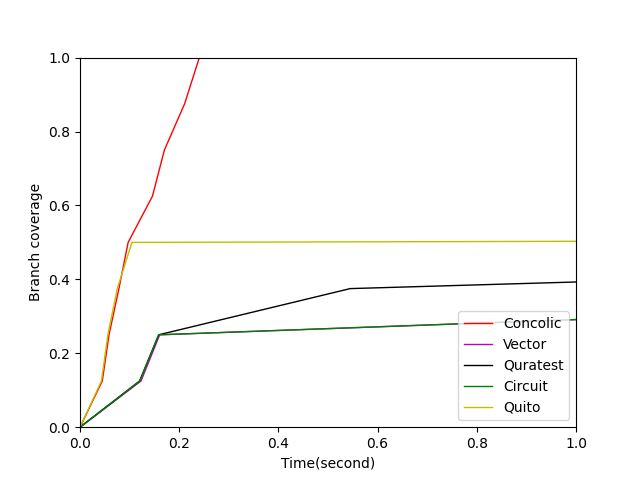}\label{fig: 1_M}}  
    \subfloat[1\_Large]{\includegraphics[width=0.33\hsize ]{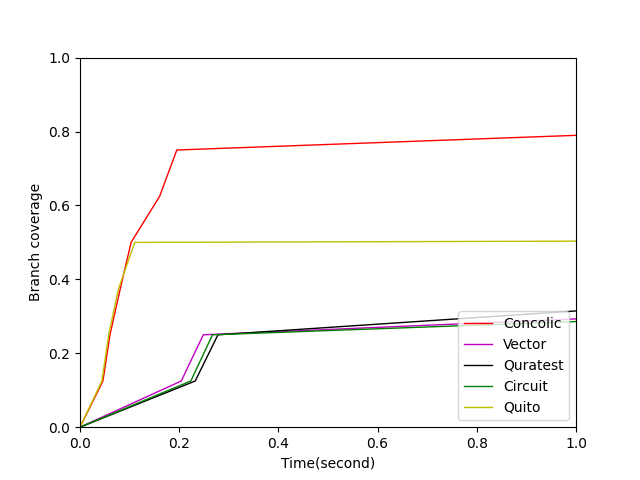}\label{fig: 1_L}}  
    
    \vspace{-0.4cm}
    \subfloat[2\_Small]{\includegraphics[width=0.33\hsize ]{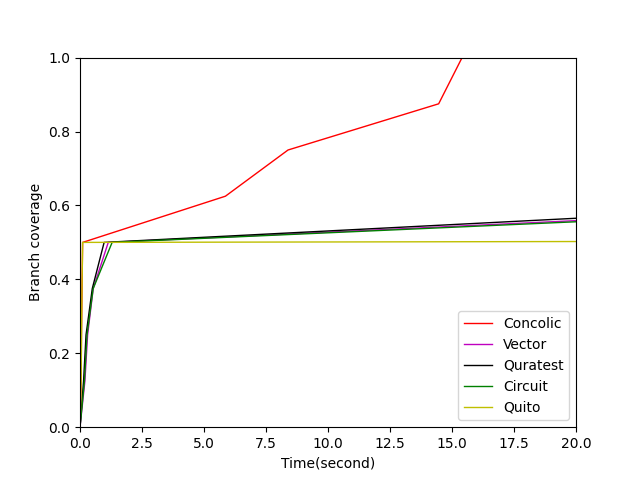}\label{fig: 2_S}}  
    \subfloat[2\_Medium]{\includegraphics[width=0.33\hsize ]{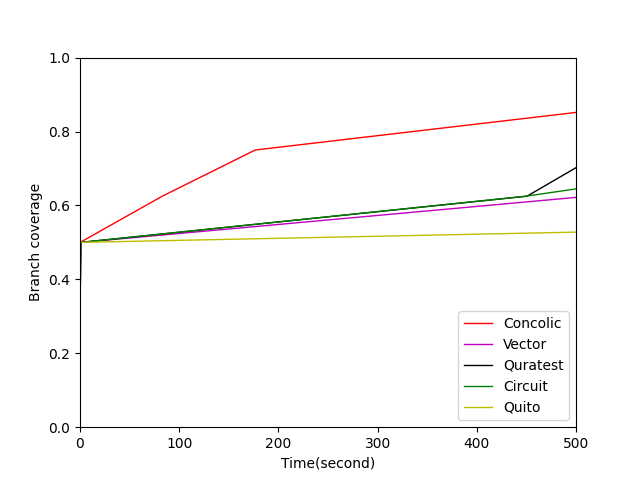}\label{fig: 2_M}}  
    \subfloat[2\_Large]{\includegraphics[width=0.33\hsize ]{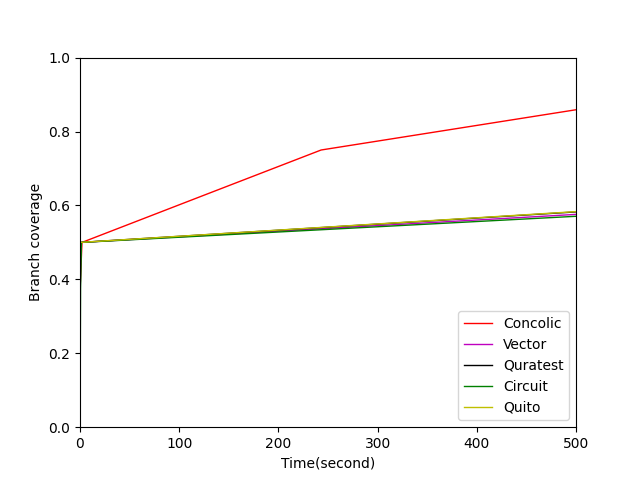}\label{fig: 2_L}}  

    \vspace{-0.4cm}
    \subfloat[3\_Small]{\includegraphics[width=0.33\hsize ]{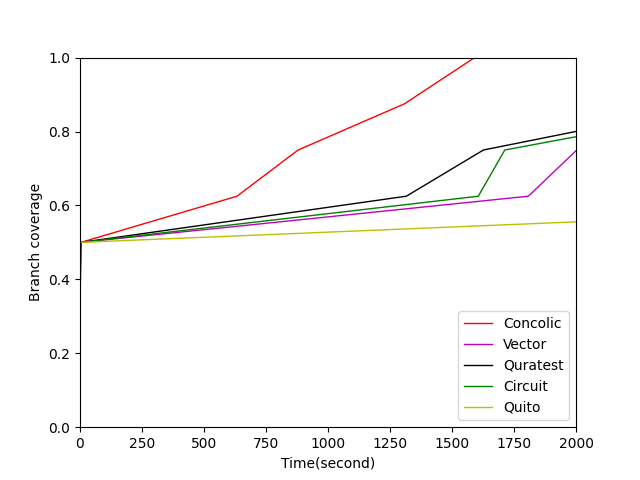}\label{fig: 3_S}}  
    \subfloat[3\_Medium]{\includegraphics[width=0.33\hsize ]{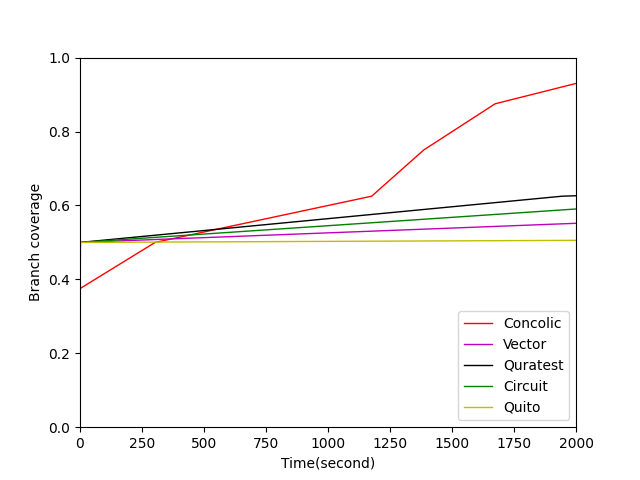}\label{fig: 3_M}}  
    \subfloat[3\_Large]{\includegraphics[width=0.33\hsize ]{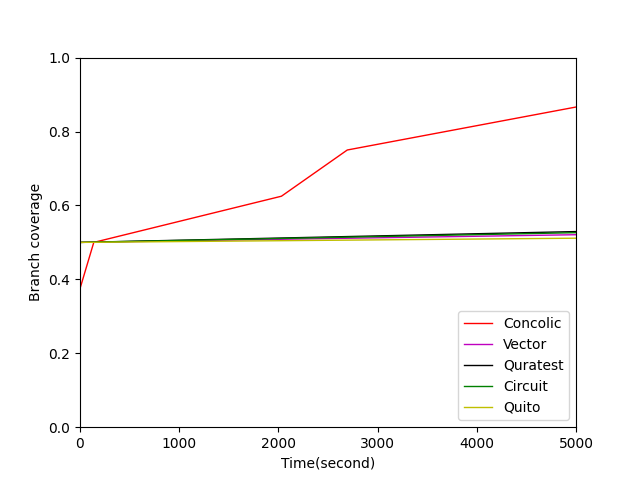}\label{fig: 3_L}}  

    \vspace{-0.4cm}
    \subfloat[4\_Small]{\includegraphics[width=0.33\hsize ]{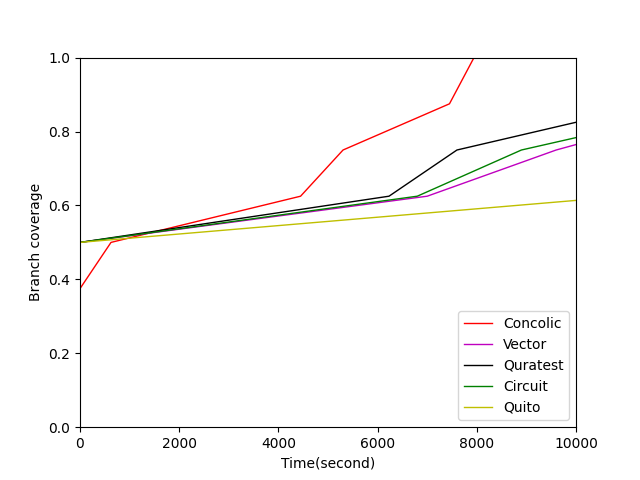}\label{fig: 4_S}}  
    \subfloat[4\_Medium]{\includegraphics[width=0.33\hsize ]{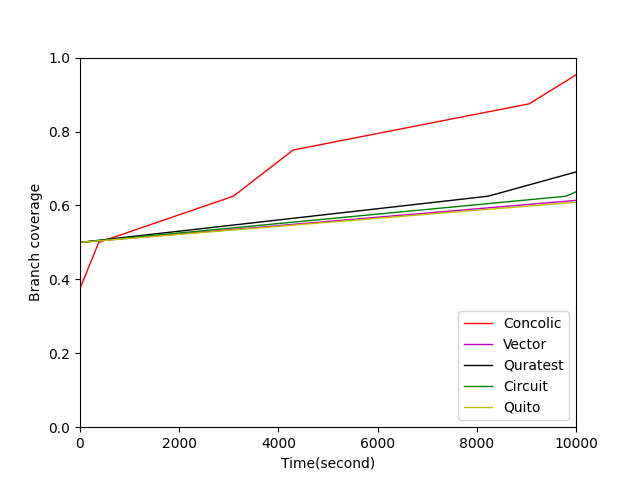}\label{fig: 4_M}}  
    \subfloat[4\_Large]{\includegraphics[width=0.33\hsize ]{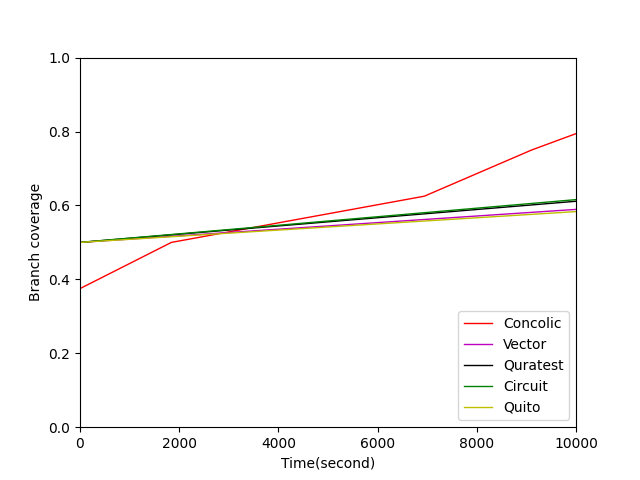}\label{fig: 4_L}}  
    \caption{Efficiency of four methods in improving branch coverage.}
    \label{fig: effective}
\end{figure*}

\textcolor{black}{In \textbf{RQ2}, we evaluate the efficiency of our method in generating effective test samples by examining the rate at which four methods improve branch coverage. We set a maximum execution time of 60,000 seconds. We record the time points at which each method increases coverage while handling different programs. The results are shown in \autoref{fig: effective}.}

\textcolor{black}{The results indicate that \textbf{Concolic} significantly outperforms the other methods in generating effective input samples and improving branch coverage efficiency, particularly with quantum programs of smaller sizes. }
\textcolor{black}{This advantage stems primarily from the fact that for quantum programs with almost infinite input space, directed generation methods eliminate test case redundancy and repetitive exploration, thus improving the efficiency of exploring program behaviors.}
\textcolor{black}{Additionally, as the number of qubits and operations increases, the time for a single concrete execution increases significantly, making the experimental time for all methods increase along with it. Due to computational pressure on the SMT solver, the efficiency of \textbf{Concolic} behaves more sensitively to variations in program size. Thus, with the refinement of quantum constraint solvers, concolic testing will perform more efficiently in quantum programs.}



\begin{tcolorbox}[size=title,rightrule=1mm, leftrule=1mm, toprule=0mm, bottomrule=0mm, arc=0pt,colback=gray!5,colframe=bleudefrance!75!black,breakable]
{ \textbf{Answer to RQ2: } 
In terms of efficiency, when facing quantum programs with insufficiently large program size, \textbf{Concolic} shows \textcolor{black}{significant} advantages in efficient test case generation and improving branch coverage. As the program size increases, the computation time of the quantum constraint solver increases with the program size, causing a loss of efficiency of \textbf{Concolic}. 
}
\end{tcolorbox}

\subsection{Quality Performance}

\textcolor{black}{In \textbf{RQ3}, we will discuss the quality performance described in Section \ref{quantum_condtion}, which is used to analyze the quality of the test cases generated by our framework. Many quantum properties exist in quantum states, some of which have been analyzed in QuraTest~\cite{ye2023quratest} to describe the diversity of test cases. Our framework is to generate test cases directly for some specific requirements. However, it is difficult to characterize directivity by using diversity metrics. Therefore, we consider the following formula:}

\vspace*{1mm}
\begin{center}
$Quality =  {\textstyle \sum_{x \in S}}  \left | |\left \langle x  | \phi  \right \rangle|^2  - p_x \right |  $    
\end{center}
\vspace*{1mm}

\textcolor{black}{where $S$ is the set of all potential observations, $\ket{\phi}$ is the quantum state obtained by the test case after the quantum program. $p_x$ is probability corresponding to the observation $x$ in the specific requirements. }

\textcolor{black}{By comparing the difference between the two probability distributions, the $Quality$ metric evaluates the gap between the test case after the execution of the quantum program and the expected requirements. $Quality$ value is closer to 0, indicating the better quality performance of the test case.}

\begin{table}[htb]
\footnotesize
\caption{RQ3 - The quality performance experiment results of the four methods.}
\label{tbl:RQ3-quality}
\centering
\setlength{\tabcolsep}{5pt}
\renewcommand{\arraystretch}{1}
\begin{tabular}{l|ccc|ccc|ccc|ccc}
\hline
          & 1\_S& 1\_M  & 1\_L & 2\_S& 2\_M  & 2\_L & 3\_S& 3\_M & 3\_L &
          4\_S& 4\_M  & 4\_L
          \\ \hline
Circuit &  0.596 & 0.599 & 0.599 & 0.828 & 0.809 & 0.834 & 0.887 & 0.852 & 0.885 &0.927 &0.901& 0.918\\
Vector &   0.559 & 0.565 & 0.560 & 0.752 & 0.741 & 0.777 & 0.854 & 0.844 & 0.881 &0.887& 0.923& 0.897\\
Quito &   0.632 & 0.585 & 0.618 & 0.817 & 0.773  & 0.797 & 0.882 & 0.876 & 0.871 &0.928& 0.936& 0.881\\
Quratest & 0.579 & 0.561 & 0.562 & 0.783 & 0.747 & 0.793 & 0.869 & 0.842 & 0.890 & 0.897& 0.901& 0.913\\
Concolic & \textbf{0.037}& \textbf{0.043} & \textbf{0.055}& \textbf{0.093}& \textbf{0.114} & \textbf{0.126}& \textbf{0.117} & \textbf{0.133}& \textbf{0.150} & \textbf{0.164} &\textbf{0.185} & \textbf{0.191}\\
\end{tabular}
\end{table}

The experiment results are in \autoref{tbl:RQ3-quality}. 
Based on the $Quality$ values, which are the average of all test cases, we find that the quality of the input samples generated by \textbf{Concolic} is better than the other four methods for all the experimental groups of quantum programs. 
\textcolor{black}{This means that the test cases generated by \textbf{Concolic} can satisfy the expected probability requirements after the quantum program. In addition, as the number of qubits increases, it becomes more difficult for the other methods to efficiently generate high-quality test cases when faced with the problem of inflated input space. \textbf{Concolic}, which is limited by the arithmetic precision of the quantum constraint solver, exhibits the problem of quality degradation with increasing program size.}


\begin{tcolorbox}[size=title,rightrule=1mm, leftrule=1mm, toprule=0mm, bottomrule=0mm, arc=0pt,colback=gray!5,colframe=bleudefrance!75!black,breakable]
{ \textbf{Answer to RQ3: } 
\textcolor{black}{In terms of quality performance, 
\textbf{Concolic} is capable of generating quantum input samples of higher quality. The quality of test cases generated by \textbf{Concolic} is affected by the computational pressure of the quantum constraint solver.
}}
\end{tcolorbox}

\subsection{Bug Finding}
\textcolor{black}{To answer \textbf{RQ4}, we evaluate the capability of test cases generated by five methods in detecting bugs in quantum programs. Our experiments are based on Bugs4Q~\cite{bugs4q}, a dataset containing 42 real-world buggy quantum programs written in Qiskit. We perform multiple random partitions of the buggy quantum programs and add quantum and classical conditional statements to them. Then, we classify them according to the eight bug patterns defined in QChecker~\cite{10189127}. Considering the execution time, we set the number of qubits for the buggy programs to 3 (this parameter is arbitrary). We used the four methods to generate test suites for the programs. For each buggy program, we generated 10 test suites, with each suite containing 30 test cases.}



\textcolor{black}{The experimental results are presented in \autoref{tb2:bug-finding}. This table records the probability that the test suites generated by the four methods detect bugs in programs classified under eight different buggy program types.}

\begin{table}[htb]
\centering
\footnotesize
\caption{RQ4 - \textcolor{black}{Eight bug patterns and capability of the four methods in terms of bug finding}.}
\label{tb2:bug-finding}
\setlength{\tabcolsep}{5pt}
\renewcommand{\arraystretch}{1}
\begin{tabular}{c|l|ccccc}
\hline
Bug Patterns & Detail Descriptions &Circuit & Vector & Quito & Quratest & Concolic \\ \hline
CE &   Call error       & 61.3\%        & 56.2\%      & 55.2\%  & 63.1\%        &\textbf{85.3\%}       \\
IG &   Incorrect uses of quantum gates       & 62.5\%       & 55.9\% & 54.4\%     & 63.4\%        &\textbf{\ 81.1\% }       \\
MI &  Measurement related issue      & 0\%        & 0\%       & 0\%   &0\%     &\textbf{\ 67.3\% }      \\
IS &   Incorrect initial state      & 57.6\%       & 54.7\%   &  55.2\%   & 64.2\%        &\textbf{\ 85.3\% }        \\
PE &    Parameter error      & 55.3\%        & 57.0\%   & 56.3\%    & 60.6\%       &\textbf{\ 82.5\% }       \\
CM &  Command misuse       & 59.1\%        & 56.3\%   & 54.2\%    & 64.5\%       &\textbf{\ 83.2\% }        \\
QE &   QASM error      & 57.4\%       & 57.5\%      & 56.5\% & 68.1\%       &\textbf{\ 81.1\% }        \\
DO &   Discarded orders      & 46.5\%       & 39.6\%    & 41.2\%   & 51.3\%       &\textbf{82.5\%}         \\ \hline
\end{tabular}
\end{table}

\textcolor{black}{Based on the experimental results, we find that \textbf{Concolic} outperforms the other methods. For most real-world buggy programs, bugs can be found by dynamic execution. Since the test suites generated by \textbf{Concolic} are direct, they can easily trigger the bugs in the program, demonstrating a high bug-finding capability in many types of buggy quantum programs. It is worth noting that for buggy programs of type MI (measurement-related issues), such bugs can cause quantum programs to fail to meet expected requirements and compromise quantum properties. However, since this type of bug does not violate programming logic or quantum computation, it is impossible to detect the problem by simply executing the program. Testing the program with the generated test suites will produce only unexecutable branches, and no errors can be reported. \textbf{Concolic} will return "unsat branch" for such bugs, indicating potential dead code and measurement misuse issues in the program.}
\textcolor{black}{Since the \emph{dReal} solver can handle the linear expressions generated by quantum operations and quadratic expressions generated by \emph{measure()}, "unsat" false alarms in the face of complex expressions are rarely encountered.}
\textcolor{black}{Notably, \textbf{Concolic} will only output the "unsat branch" statement in the quantum constraint part. This statement occurs primarily when the superposition property of the qubit is destroyed using \emph{measure()} before reaching the constraint during execution, but the constraint requires that this qubit be still in superposition.}

\begin{tcolorbox}[size=title,rightrule=1mm, leftrule=1mm, toprule=0mm, bottomrule=0mm, arc=0pt,colback=gray!5,colframe=bleudefrance!75!black,breakable]
{ \textbf{Answer to RQ4: } 
\textcolor{black}{
In terms of bug-finding capability, 
\textbf{Concolic} is effective in generating test suites that trigger program bugs compared to other approaches. Additionally, \textbf{Concolic} can detect bugs caused by the misuse of quantum measurements. However, the other three methods are difficult to detect effectively.
}}
\end{tcolorbox}

\section{Threats to Validity}\label{sec:threats}

\textcolor{black}{We recognize that randomness plays a significant role in the generation and execution of tests. To mitigate this factor, we repeated each experimental configuration ten times and presented the average results in the paper. Additionally, the selection of quantum programs for testing could introduce bias. We intentionally designed benchmarks with four different branching structures to counteract this. Furthermore, all constraints and quantum operations used in the Qiskit programs during our experiments were generated randomly, aiming to reduce selection bias.}

\textcolor{black}{A further potential threat involves the metrics used in RQ3. Currently, there is yet to be a universally accepted definition of the quality of input samples generated for quantum programs, which complicates the evaluation of this attribute. Consequently, the metrics we have proposed are intended as a means of providing a measurable assessment of this elusive quantum characteristic.}

\section{Related Work}\label{sec:related}

\textcolor{black}{Recently, quantum programs have gained significant attention, and numerous studies have emerged to analyze their behavior. Two such studies, by Ali \textit{et al.}~\cite{ali2021assessing} and Wang \textit{et al.}~\cite{quito}, have introduced the first coverage criterion, focusing on the inputs and outputs of quantum programs. Furthermore, Zhao \textit{ et al.}~\cite{bugs4q} have reported a data set of 42 bugs found in Qiskit quantum programs and have also summarized eight bug patterns in \cite{bug-detection-1}. These studies emphasize the importance of testing quantum programs to ensure their reliability.}

\textcolor{black}{Automated test case generation is a crucial aspect of software engineering that ensures software quality and reliability. Several techniques have been proposed to automate the generation of test cases for quantum programs, such as fuzzing testing~\cite{wang2018quanfuzz,10.1007/978-3-030-98012-2_30,91f691381d2f465382430e1c317c5301}, mutation testing~\cite{mutateali, Fortunato2022, muskit, FortunatoICSE2022}, and metamorphic testing~\cite{Paltenghi_2023,9814906}. Wang ~\cite{wang2021generating, wang2022qusbt} has developed search-based test case generation, which has been developed by Wang~\cite{wang2021generating,wang2022qusbt} for quantum programs to detect bugs by generating binary test cases. Long and Zhao~\cite{long2022testing,long2024testing} introduced a systematic testing framework that includes unit and integration testing. On the other hand, Ye \emph{et al.}~\cite{ye2023quratest} proposed QuraTest, which integrates quantum properties into test case generation. However, these frameworks often produce numerous low-quality, redundant samples that do not enhance the branch coverage of the program. When dealing with quantum programs with complex branching structures, it is challenging for these methods to uncover more hidden erroneous statements. In contrast, our quantum concolic testing framework uses a branch-and-bound exploration and path constraint-solving approach for input sample generation. It directly targets specific branches, making it more effective in identifying bugs.}

\textcolor{black}{Given the value of quantum resources, the benefits of symbolic execution have also captured the attention of researchers, leading to significant contributions. SymQV~\cite{bauer2023symqv} was introduced as the first quantum symbolic execution framework and has been applied to the automated verification of quantum programs. Recently, Fang \emph{et al.}~\cite{fang2023symbolic} proposed a symbolic execution framework, QSE, for quantum error correction programs, providing a soundness proof for their method. However, these frameworks require formal verification expertise and are primarily designed to verify quantum algorithms rather than test arbitrary quantum programs.}

\textcolor{black}{Concolic testing, a variation of symbolic execution that combines symbolic and concrete execution, was first introduced by Godefroid \emph{et al.}~\cite{godefroid2005dart} through the DART tool for automatically testing software. DART generates constraints for execution paths based on specific inputs and explores new paths to generate test cases and increase test coverage. Building on this concept, Sen \emph{et al.}~\cite{sen2005cute} developed a concolic unit testing engine for C programs, and Sun \emph{et al.}~\cite{sun2018concolic} adapted it for deep neural networks. Majumdar \emph{et al.} improved the approach to concolic testing by integrating the strengths of random search and concolic testing. Nevertheless, these frameworks are not directly applicable to quantum programs.}

\section{Conclusion}\label{sec:conclusion}
In this work, we propose the first concolic testing framework for quantum programs.
\textcolor{black}{We first attempt to address the challenges in generating quantum constraints for each execution path and devise three constraint generation methods to address the shortcomings of measurement-based quantum conditions. Using our method, we evaluated branch coverage, the efficiency of generating input samples, the quality of input samples, and the capability of detecting quantum bugs. Our in-depth evaluation demonstrates that our proposed testing techniques can efficiently generate high-quality test samples and detect bugs in quantum features. We also explore the impact of changes in qubit count and program size on the method for generating input samples for quantum programs.}
Our work provides a valuable contribution to quantum software testing and can potentially improve the quality and reliability of quantum programs.

\textcolor{black}{Quantum concolic testing has shown promising results in improving branch coverage for quantum programs. However, it heavily depends on the efficiency of quantum constraint solvers. Therefore, optimizing these quantum constraints can be a potential area for further improvement to reduce computational demands on the solver. We also plan to extend our quantum concolic testing framework to support additional quantum programming language platforms.}

\section{Data Availability}
Our framework is publicly available at \url{https://github.com/Xzore19/quantum_concolic}.

\section*{Acknowledgments}
This work was supported by JST SPRING Grant No.\ JPMJSP2136, JST BOOST Grant No.\ JPMJBS2406, and JSPS KAKENHI Grants No.\ JP23K11049, No.\ JP23H03372, and No.\ JP24K14908.

\appendix
\section{Quantum Constraint and SMT file.}
\label{smt}

We will use one of the execution paths in example \autoref{fig:example} to show the structure of the quantum constraints and the corresponding SMT file.
We consider one of the execution paths with the most quantum operations: the results of lines (14, 16, and 21) are (True, True, False), which are generated by our framework, is 

\begin{center}
    
$qc\_constraint: (in\{['x(1)', 'z(1)', 'h(1)']\} \ qc\#[0.+0.j, 0.49875801-0.50370759j, 0.+0.j, 0.49375883-0.50370759j]|, ([1] [1], ['0']))(False)$

\end{center}

\noindent
where $in$ is a flag for \emph{measure()}. $['x(1)', 'z(1)', 'h(1)']$ are all quantum operations in the current execution paths. $[0.+0.j, 0.49875801-0.50370759j, 0.+0.j, 0.49375883-0.50370759j]$ is the test case for this concrete execution to generate (True, True, False). $([1] [1], ['0'])$ is the detail of $measure()$ and the result is $(False)$.

For this current execution path, our next direction of exploration is (True, True, True). The corresponding SMT file consists of the following five modules:

\begin{enumerate}[leftmargin=2em] 
\setlength{\itemsep}{2pt}

\item \textbf{Define variables}:

(declare-fun psi\_0\_3.i () Real) (declare-fun psi\_0\_3.r () Real) (declare-fun psi\_0\_2.i () Real) (declare-fun psi\_0\_2.r () Real)
(declare-fun psi\_0\_1.i () Real) (declare-fun psi\_0\_1.r () Real) (declare-fun psi\_0\_0.i () Real) (declare-fun psi\_0\_0.r () Real) ......

\item \textbf{Basic constraint for qubits}:

(assert (= (+ 0.0      (\^{} psi\_0\_0.r 2.0)      (\^{} psi\_0\_0.i 2.0)      (\^{} psi\_0\_1.r 2.0)      (\^{} psi\_0\_1.i 2.0)      
(\^{} psi\_0\_2.r 2.0)      (\^{} psi\_0\_2.i 2.0)      (\^{} psi\_0\_3.r 2.0)      (\^{} psi\_0\_3.i 2.0))   1.0))

\item \textbf{Quantum operations} (for example, X gate):

(assert (let ((a!1 (= (+ 0.0 (- (* 1.0 psi\_0\_2.r) (* 0.0 psi\_0\_2.i))) psi\_1\_0.r))      (a!2 (= (+ 0.0 (- (* 1.0 psi\_0\_3.r) (* 0.0 psi\_0\_3.i))) psi\_1\_1.r))      
(a!3 (= (+ 0.0 (- (* 1.0 psi\_0\_0.r) (* 0.0 psi\_0\_0.i))) psi\_1\_2.r))      (a!4 (= (+ 0.0 (- (* 1.0 psi\_0\_1.r) (* 0.0 psi\_0\_1.i))) psi\_1\_3.r)))  
(and a!1       (= (+ 0.0 (* 1.0 psi\_0\_2.i) (* 0.0 psi\_0\_2.r)) psi\_1\_0.i)       a!2       (= (+ 0.0 (* 1.0 psi\_0\_3.i) (* 0.0 psi\_0\_3.r)) psi\_1\_1.i)       
a!3       (= (+ 0.0 (* 1.0 psi\_0\_0.i) (* 0.0 psi\_0\_0.r)) psi\_1\_2.i)       a!4       (= (+ 0.0 (* 1.0 psi\_0\_1.i) (* 0.0 psi\_0\_1.r)) psi\_1\_3.i))))

\item \textbf{Quantum control statements}:

(assert (or (and (= psi\_3\_1.r 0.0)         (= psi\_3\_1.i 0.0)         (= psi\_3\_3.r 0.0)         (= psi\_3\_3.i 0.0))))

\item \textbf{SMT statements}:

(check-sat)

(get-model)

\end{enumerate}

By solving this SMT file, we obtain a new result $[(0.4474683797931982-0.4996736486934912j), 0j, 
\\(0.5398380316790629-0.507317399153358j), 0j]$ for the next execution and exploration.

\bibliographystyle{ACM-Reference-Format}
\bibliography{ref}

\end{document}

%% file: macros.tex
\usepackage{times}
\usepackage{graphicx}
\usepackage{epsf}
\usepackage{verbatim}
\usepackage{url}
\usepackage{color}
\usepackage{alltt}

\newcommand{\Comment}[1]{}

\newcommand{\SmallSpace}{\vspace*{-1.4ex}}

